\documentclass[conference]{IEEEtran}
\IEEEoverridecommandlockouts
\usepackage{cite}
\usepackage{amsmath,amsfonts}
\usepackage{algorithmic}
\usepackage{array}
\usepackage[caption=false,font=normalsize,labelfont=sf,textfont=sf]{subfig}
\usepackage{textcomp}
\usepackage{stfloats}
\usepackage{url}
\usepackage{verbatim}
\usepackage{graphicx}

\usepackage{amsfonts}
\usepackage{dsfont}

\usepackage{array,longtable}

\usepackage{amssymb}

\usepackage{amsmath}

\usepackage{amsthm}

\usepackage{amsmath}

\usepackage{graphicx}

\usepackage{epstopdf}

\usepackage[hidelinks]{hyperref}

\usepackage{epsfig}
\usepackage{mdframed}

\usepackage{float}

\usepackage{array,longtable}

\usepackage[singlelinecheck=off]{caption}

\usepackage{url}

\PassOptionsToPackage{hyphens}{url}\usepackage{hyperref}

\usepackage{lipsum}
\newtheorem{definition}{Definition}[]

\newtheorem{theorem}{Theorem}[]

\usepackage{mathtools}

\usepackage{xtab,booktabs}

\newcommand{\kw}[1]{{\color{blue}\texttt#1}}

\newcommand{\holl}{\textsf{HOL Light}}

\usepackage{listings}

\usepackage{graphicx}
\usepackage{amsfonts}
\usepackage{dsfont}
\usepackage{bbm}

\def\BibTeX{{\rm B\kern-.05em{\sc i\kern-.025em b}\kern-.08em
    T\kern-.1667em\lower.7ex\hbox{E}\kern-.125emX}}
\usepackage{balance}

\begin{document}
\title{Formalization of Biological Circuit Block Diagrams for formally analyzing Biomedical Control Systems in pHRI Applications \thanks{This work was supported and funded by Kuwait University, Research Project No. (EO 01/23).\\ *Corresponding Author}}


\author{\IEEEauthorblockN{1\textsuperscript{st} Adnan Rashid}
\IEEEauthorblockA{\textit{School
of Electrical Engg. and Computer Science} \\
\textit{National University of Sciences and Technology (NUST)}\\
Islamabad, Pakistan \\
adnan.rashid@seecs.edu.pk}
\and
\IEEEauthorblockN{2\textsuperscript{nd} Saed Abed*}
\IEEEauthorblockA{\textit{Computer Engg. Depertment, College of Engg. and Petroleum} \\
\textit{Kuwait University}\\
Kuwait City, Kuwait \\
s.abed@ku.edu.kw}
\and
\IEEEauthorblockN{3\textsuperscript{rd} Osman Hasan}
\IEEEauthorblockA{\textit{School
of Electrical Engg. and Computer Science} \\
\textit{National University of Sciences and Technology (NUST)}\\
Islamabad, Pakistan\\
osman.hasan@seecs.edu.pk}}


\maketitle

\begin{abstract}
 The control of Biomedical Systems in Physical Human-Robot Interaction (pHRI) plays a pivotal role in achieving the desired behavior by ensuring the intended transfer function and stability of subsystems within the overall system. Traditionally, the control aspects of biomedical systems have been analyzed using manual proofs and computer based analysis tools. However, these approaches provide inaccurate results due to human error in manual proofs and unverified algorithms and round-off errors in computer-based tools. We argue using Interactive reasoning, or frequently called theorem proving, to analyze control systems of biomedical engineering applications, specifically in the context of Physical Human-Robot Interaction (pHRI). Our methodology involves constructing mathematical models of the control components using Higher-order Logic (HOL) and analyzing them through deductive reasoning in the HOL Light theorem prover. We propose to model these control systems in terms of their block diagram representations, which in turn utlize the corresponding differential equations and their transfer function-based representation using the Laplace Transform (LT). These formally represented block diagrams are then analyzed through logical reasoning in the trusted environment of a theorem prover to ensure the correctness of the results. For illustration, we present a real-world case study by analyzing the control system of the utrafilteration dialysis process.

\end{abstract}



\section{Introduction} \label{SEC:intro}

Biomedical systems encompass a wide range of technologies and processes used in healthcare, including medical devices, diagnostic tools, monitoring equipment, and life-supporting machines, as well as processes such as dialysis or drug delivery systems. These systems interact directly with the human body and are often responsible for life-sustaining functions, making their reliability and safety paramount. Given that even minor malfunctions or inaccuracies in these systems can lead to severe consequences, such as incorrect diagnoses, improper treatment, or even loss of life, ensuring their correctness is critical. Biomedical systems must adhere to stringent safety and performance standards to minimize risks to patients. The complexity of these systems, often involving continuous data processing, real-time decision-making, and interactions with biological processes, further emphasizes the need for rigorous verification. Any failure in the design or operation of biomedical systems can compromise patient safety, highlighting the importance of thorough testing, validation, and formal verification methods to guarantee their dependability and effectiveness in clinical settings.

Nowadays principles from engineering, mathematics, and physics have been extensively employed to analyze biomedical systems, such as prosthetic eyes, artificial limbs, pacemakers, dental implants, and dialysis machines. In these systems, control components are crucial for achieving desired behaviors by regulating the functioning of different modules, ensuring smooth operation of the entire system. For instance, automatic anesthesia controllers administer drugs to patients to prevent overdosing, and automatic control of hemodialysis offers better treatment for end-stage renal disease patients. Similarly, control components in medical robots assist in rehabilitating disabled patients.

Analyzing these biomedical systems requires modeling their dynamics to understand interactions between different system modules. These dynamic behaviors are typically represented using differential equations that capture relationships between various parameters, such as input signals and corresponding system responses. The Laplace Transform (LT) is subsequently applied to transform these differential equations from the time domain into their frequency-domain equivalents. These models form the basis for performing Transfer Function (TF) and stability analyses of the control components in biomedical systems.

Historically, the control aspects of biomedical systems have been analyzed using manual proofs and computer-based symbolic and numerical methods~\cite{fernandez2018automatic}. However, manual approaches are prone to human error, particularly when handling large and complex systems. There's also a risk of overlooking critical assumptions necessary for accurate mathematical analysis, potentially leading to erroneous results. Computer-based methods rely on tools like Maple, MATLAB, and Mathematica, which may contain unverified algorithms or produce incorrect results due to unproven symbolic procedures or numerical approximations.

The increasing use of robotic systems in biomedical engineering, especially in applications involving Physical Human-Robot Interaction (pHRI), necessitates rigorous verification techniques to ensure safety and reliability. pHRI systems, which involve direct or indirect physical contact between humans and robots, are especially prone to safety risks. Therefore, ensuring the correct operation of control systems in these applications is crucial. Therefore, given the safety-critical nature of these biomedical systems, the above-mentioned conventional approaches cannot be relied upon.

Formal verification techniques~\cite{hasan2015formal} are computer-assisted mathematical methods commonly used to rigorously model and analyze complex systems~\cite{beillahi2015formal,siddique2013formal}. One of the most widely adopted formal methods is Interactive Theorem Proving (ITP)\cite{hasan2015formal}, which employs suitable logics—such as propositional, predicate, or higher-order logic—to build mathematical models of systems. These models are then verified against their intended specifications through deduction and reasoning within a theorem prover, ensuring the soundness and completeness of the analysis. While ITP has been applied to Laplace Transform-based analyses in various engineering domains, including automated vehicle platoons\cite{rashid2018formal}, linear analog filters~\cite{taqdees2017formally}, unmanned submersible vehicles~\cite{rashid2017formal}, image processing filters~\cite{adnanimage}, cyber physical systems~\cite{adnancyber}, to our knowledge, it has not yet been utilized for formal verification of control systems in biomedical engineering, particularly in pHRI.

This paper introduces a formal analysis approach based on interactive theorem proving for analyzing control aspects in biomedical systems. Specifically, we propose to develop formal models capturing the dynamics of control components using differential equations in higher-order logic. We then  transform these equations into their frequency-domain representations using the Laplace Transform. These transformed representations are subsequently used to conduct transfer function (TF) and stability analyses of the control components within the HOL Light theorem prover.
In this step, we require the block diagrams representations that
allow us to establish a relationship of time-domain dynamics of these systems with
their corresponding frequency domain representations. The block diagrams facilitate the breakdown of complex biomedical control systems into simpler, analyzable components, enabling systematic verification and control design.

We introduced a formalization approach using higher-order logic for block diagram representations of biological circuits in \cite{rashid2020syntheticbiology}. This framework was applied to rigorously analyze protein activation and repression, as well as autoactivation, along with phase lead and lag controllers. These components are commonly employed in cancer cell identification and multi-input receptors for accurate disease diagnosis. We developed a similar formalization of block diagrams in this current paper but the focus of this block diagram formalization is for analyzing the control aspects of biomedical systems. So despite being block diagram representations the context is totally disjoint. We chose HOL Light for our formalization because of its comprehensive libraries for multivariate calculus and LT. For illustration purposes, we formally verify the
Ultrafiltration dialysis process, which is a process of removing a volume
of excess fluid from a patient by separating the small particles
and macromolecules from the body fluid water.


\section{Preliminaries} \label{SEC:prelim}


\subsection{\holl}\label{SUBSEC:tp_hol_light_tp}
Theorem proving is a formal verification method used to verify mathematical theorems through computer programs known as theorem provers. This approach involves creating a formal model of a system and verifying its properties using higher-order logic, which offers greater expressiveness by allowing additional quantifiers. The undecidable nature of the underlying higher-order logic leads to the involvement of significant human interaction for proof construction, making these tools interactive theorem provers. They are particularly suitable for mathematical analysis based on multivariate calculus. HOL Light is a widely-used interactive theorem prover, implemented in OCaml~\cite{Ocaml_ref}, and designed to automate mathematical proofs. It has a minimal trusted core (about 400 lines of OCaml code), which has been verified through the CakeML project~\cite{harrison2006towards}.

\subsection{Multivariable Calculus Theories of \holl}
\holl~offers extensive support for system analysis through its multivariable calculus theories, like integral~\footnote{\url{https://github.com/jrh13/hol-light/blob/master/Multivariate/integration.ml}}, derivatives~\footnote{\url{https://github.com/jrh13/hol-light/blob/master/Multivariate/derivatives.ml}}, vectors~\footnote{\url{https://github.com/jrh13/hol-light/blob/master/Multivariate/vectors.ml}}, transcendental~\footnote{\url{https://github.com/jrh13/hol-light/blob/master/Multivariate/transcendentals.ml}} and topology~\footnote{\url{https://github.com/jrh13/hol-light/blob/master/Multivariate/topology.ml}}.
All theorems in these theories are formally verified for generic functions, i.e., functions with data type $\mathbb{R^{N}} \to \mathbb{R^{M}}$. Some of these functions that are directly used in our proposed methodology for the formal verification of control aspects in biomedical systems are described below.

\begin{mdframed}
\begin{definition}
\label{DEF:integral_mc}
\emph{Integral of a Vector-valued Function} \\
{
\small
\textup{\texttt{
$\vdash_{def}$ $\forall$g x. \kw{integ} x g =   \\
  \hspace*{1.5cm} (@i.(g has\_integ i) x)
}}}
\end{definition}
\end{mdframed}

The above-mentioned function returns the integral of  \texttt{g} in the region of integration \texttt{x:} $\mathbb{R^{N}} \to \mathbb{B}$, which represents a vector space. The function \texttt{has\_integ} represents the integral relationship in relational form. The Hilbert choice operator \texttt{@:} $\mathbb{(\alpha \to \texttt{bool}) \to \alpha}$ provides the value of the integral, if it exists.

\begin{mdframed}
\begin{definition}
\label{DEF:derivative_mc}
\emph{Derivative of a Vector-valued Function} \\
{
\small
\textup{\texttt{
$\vdash_{def}$ $\forall$g n. \kw{vec\_der} g n =     \\
  \hspace*{0.6cm} (@g'.(g has\_vec\_der f') n)
}}}
\end{definition}
\end{mdframed}

The above function provides the differential of the function \texttt{g} at a point \texttt{n:} $\mathbb{R}^{N} \to \mathbb{B}$. It returns a vector of type $\mathbb{R^{M}}$, which represents the derivative of \texttt{g} at \texttt{n}.




\subsection{Laplace transform Theories of \holl} \label{SEC:form_lap_trans}

The LT of a vector-valued continuous-time function is expressed as~\cite{adnan2018JAL}:


\begin{equation}\label{EQ:laplace_transform}
\mathcal{L} [f(t)] = F(s) = \int_{0}^{\infty} {f(t)e^{-s t}} dt, \ \ s \in  \mathds{C}
\end{equation}


Here, $s$ represents a complex variable. Additionally, $\mathds{C}$
represents the data type of a complex number, which corresponds to the data type $\mathbb{R}^{2}$
in \holl, essentially a column matrix with two elements. The LT (Equation~(\ref{EQ:laplace_transform}))  has been formalized in \holl as detailed in \cite{adnan2018JAL}:

\begin{mdframed}
\begin{definition}
\label{DEF:laplace_transform}
\emph{Laplace Transform} \\
{
\small
\textup{\texttt{
$\vdash_{def}$ $\forall$s g. \kw{ltfm} g s =  \\
 \hspace*{0.8cm} integ \{x| \&0 $\leq$ $\underline{\texttt{x}}$\}   \\
  \hspace*{1.5cm} ($\lambda$x. cexp (--(s $\ast$ Cx $\underline{\texttt{x}}$)) $\ast$ g x)
}}}
\end{definition}
\end{mdframed}

\noindent where $\texttt{cexp} : \mathds{C} \rightarrow \mathds{C}$ models the complex-valued exponential function. Similarly, $\underline{\texttt{t}}$ represents the \holl~function \texttt{drop t} that converts a one-dimensional vector \texttt{t} to a real number, and the operator \texttt{\&} typecasts a natural number into a real number. Similarly, \texttt{{t | \&0 $\leq$ $\underline{\texttt{t}}$}} defines the positive real line, representing the integration limit from 0 to  $\infty$. In the above definition, the function \texttt{ltfm} returns the LT of \texttt{f} as per Equation~(\ref{EQ:laplace_transform}) by utilizing the vector function \texttt{integral}.

The Laplace transform exists for piecewise smooth functions as well as the functions for which the exponential order is on the positive real
line. These conditions are met if the given function is piecewise differentiable within the given interval. We formally captured this as follows~\cite{adnan2018JAL}:

\begin{mdframed}
\begin{definition}
\label{DEF:laplace_existence}
\emph{Existence of the Laplace Transform} \\
{
\small
\textup{\texttt{$\vdash_{def}$ $\forall$s g. \kw{lexst} g s $\Leftrightarrow$  \\
\hspace*{0.0cm} ($\forall$b. g pcws\_diff\_on \\
\hspace*{4.2cm}  interval [$\underline{\texttt{\&0}}$,$\underline{\texttt{b}}$]) $\wedge$  \\
\hspace*{0.0cm}  ($\exists$M a. Re s > $\underline{\texttt{a}}$ $\wedge$ eord g M a) }}}
\end{definition}
\end{mdframed}

\noindent The function \texttt{Re}  returns the real part of its input complex number. The function \texttt{eord} models the exponential-order condition to ensure that the LT exists and its formalization is detailed in~\cite{taqdees2013formalization,adnan2018JAL}:

\begin{mdframed}
\begin{definition}
\label{DEF:exp_order_condition}
\emph{Exp Order} \\
{
\small
\textup{\texttt{$\vdash_{def}$ $\forall$g M a. \kw{eord} g M a $\Leftrightarrow$  \\
\hspace*{0.0cm} \&0 < M $\wedge$   \\
\hspace*{0.0cm}   ($\forall$t. \&0 $\leq$ t $\Rightarrow$  ||g $\overline{\texttt{t}}$|| $\leq$ M $\ast$ exp ($\underline{\texttt{a}}$ $\ast$ t)) }}}
\end{definition}
\end{mdframed}

\noindent Here, $||\vec{\texttt{x}}||$ denotes the norm of the vector. Likewise, $\overline{\texttt{t}}$ refers to the \holl~function \texttt{lift t}, which converts a real number \texttt{t} into a one-dimensional vector.


We utilized the formal definitions of the Laplace transform and its existence condition to verify the commonly used properties such as linearity, time and frequency shifting, time scaling, differentiation and integration in time domain in \cite{adnan2018TC}.

\section{Formalization of Block Diagram Representations}\label{SEC:form_block_diagram_rep}

 In biomedical systems involving human-robot interactions (pHRI), biological functions are often governed by intricate interactions between various physiological components. Block diagrams help break down these systems into distinct functional units or modules, each representing specific processes, such as fluid regulation, filtration, or control mechanisms in human-robot collaboration.

We now introduce our formal definitions for the essential building blocks (subsystems) of these block diagram representations. These definitions allow for the formal modeling of block diagrams for any biomedical system or process in the s-domain and enable the determination of the transfer function from its block diagram. The proposed formalization is primarily inspired by the block diagrams used in control systems~\cite{houpis2013linear} and our previous formal modeling of block diagram representations~\cite{rashid2020syntheticbiology}.

\begin{figure}[!ht]
\centering
\scalebox{0.350}
{\hspace*{-0.4cm} \includegraphics[trim={5.0 0.4cm 5.0 0.4cm},clip]{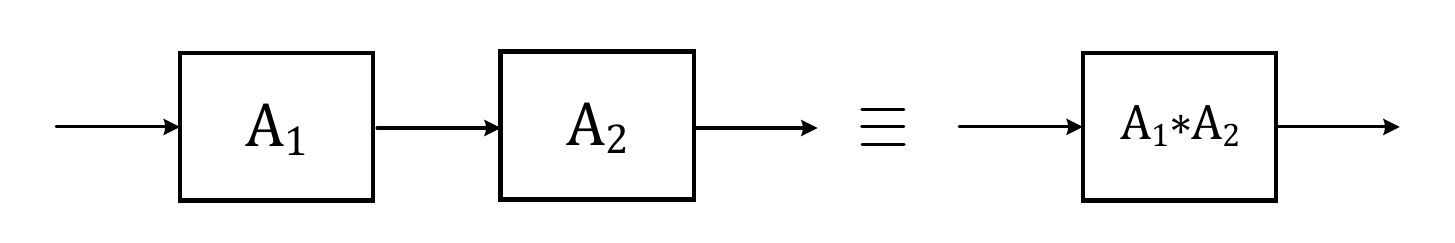}}
\caption{Series Configuration}
\label{FIG:series_rep}
\end{figure}

\textbf{Serial Configuration} The transfer function for the series connection of two components (subsystems), as shown in Figure~\ref{FIG:series_rep}, in a biomedical system, such as an ultrafiltration dialysis process—representing fluid transfer between different parts of the patient like the arm, trunk, or leg, can be expressed as the multiplication of the individual transfer functions  as shown in Figure~\ref{FIG:series_rep}. We model in a generic way ((\textit{N}) components) as follows:

\begin{mdframed}
\begin{definition}
\label{DEF:series_rep_comp}
\emph{Series Configuration} \\
{
\small
\textup{\texttt{\textsf{
$\vdash_{\mathit{def}}$ $\forall$A$\mathtt{\mathsf{_i}}$. \kw{ser} [A$\mathtt{\mathsf{_1}}$; A$\mathtt{\mathsf{_2}}$; ...; A$\mathtt{\mathsf{_N}}$] = $\mathtt{\mathsf{\prod\limits_{i = 1}^{N}}}$ A$\mathtt{\mathsf{_i}}$
}}}}
\end{definition}
\end{mdframed}

The above-mentioned function returns the overall transfer function of an arbitrary number of block connected in series, which it takes their individual transfer functions as input in the form of a list. The net transfer function is calculated by multiplying all the individual transfer functions.

\begin{figure}[!ht]
\centering
\scalebox{0.330}
{\hspace*{0.0cm} \includegraphics[trim={5.0 0.4cm 5.0 0.4cm},clip]{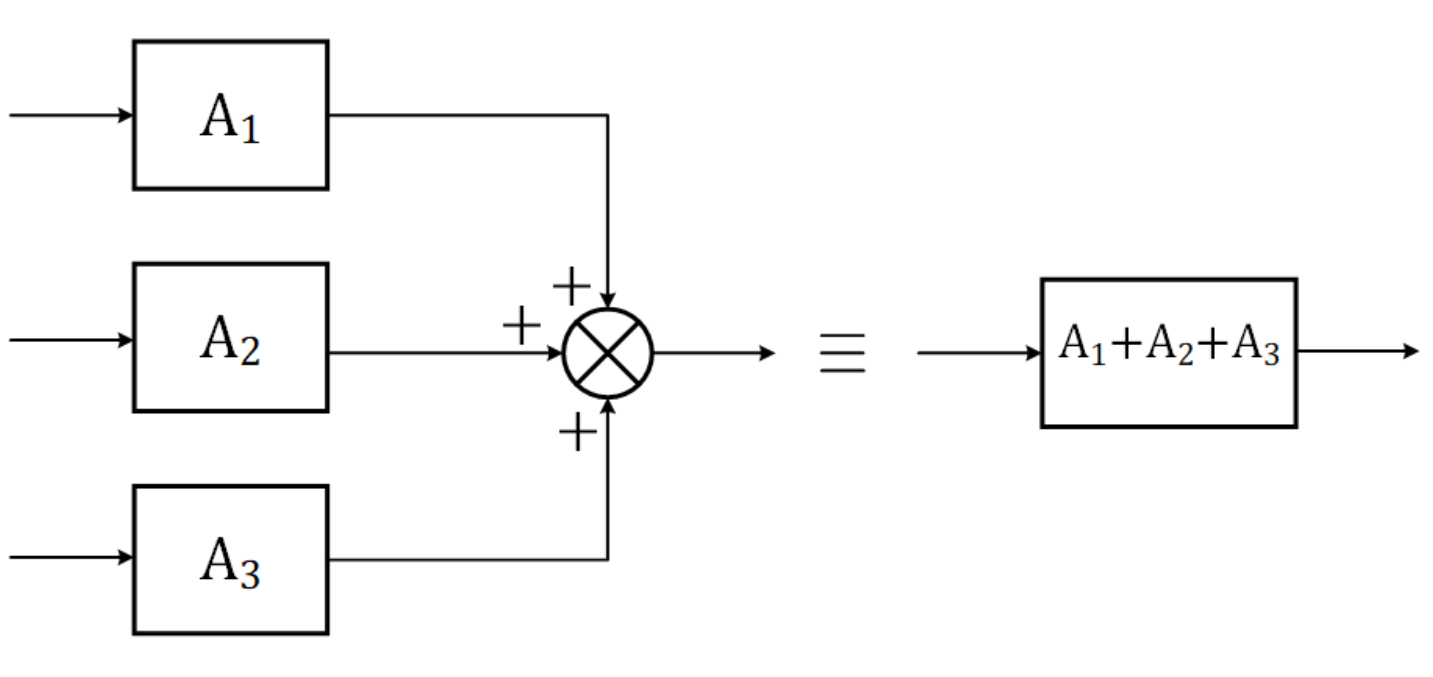}}
\caption{Summation Junction}
\label{FIG:summ_junc}
\end{figure}

\textbf{Summation Junction} It acts as an addition module, as illustrated in Figure~\ref{FIG:summ_junc}, that adds the transfer functions of all the individual components.

\begin{mdframed}
\begin{definition}
\label{DEF:summation_jun}
\emph{Summation} \\
{
\small
\textup{\texttt{\textsf{
$\vdash_{\mathit{def}}$ $\forall$A$\mathtt{\mathsf{_i}}$. \kw{summ} [A$\mathtt{\mathsf{_1}}$; A$\mathtt{\mathsf{_2}}$; ...; A$\mathtt{\mathsf{_N}}$] = $\mathtt{\mathsf{\sum\limits_{i = 1}^{N}}}$ A$\mathtt{\mathsf{_i}}$
}}}}
\end{definition}
\end{mdframed}

\begin{figure}[!ht]
\centering
\scalebox{0.330}
{\hspace*{0.0cm} \includegraphics[trim={5.0 0.4cm 5.0 0.4cm},clip]{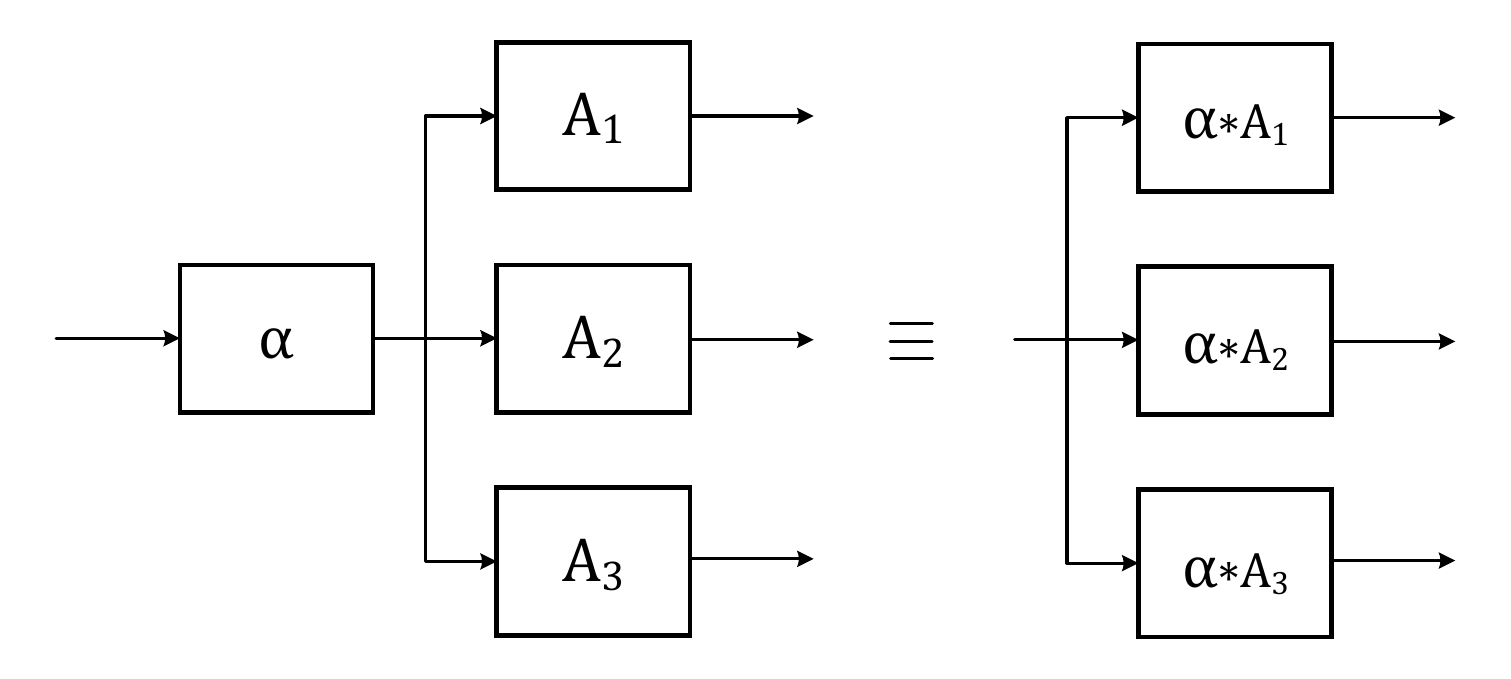}}
\caption{Pickoff Point}
\label{FIG:pickoff_point}
\end{figure}

\textbf{Pickoff} The pickoff point configuration represents the branching of a component into parallel components (Figure~\ref{FIG:pickoff_point}). We formally model this as follows:

\begin{mdframed}
\begin{definition}
\label{DEF:pickoff_point}
\emph{Pickoff} \\
{
\small
\textup{\texttt{\textsf{
$\vdash_{\mathit{def}}$ $\forall \mathtt{\mathsf{\alpha}}$ A$\mathtt{\mathsf{_i}}$. \kw{pick\_point} [A$\mathtt{\mathsf{_1}}$; A$\mathtt{\mathsf{_2}}$; ...; A$\mathtt{\mathsf{_N}}$] =  \\
 \hspace*{4.5cm} [$\mathtt{\mathsf{\alpha}}\ \ast$ A$\mathtt{\mathsf{_1}}$; $\mathtt{\mathsf{\alpha}}\ \ast$ A$\mathtt{\mathsf{_2}}$; ...; $\mathtt{\mathsf{\alpha}}\ \ast$ A$\mathtt{\mathsf{_N}}$]
}}}}
\end{definition}
\end{mdframed}

The above-mentioned function returns the equivalent transfer functions arranged as a list of components by using the transfer functions of the first and the rest of the components, represented as a list of complex numbers.

\begin{figure}[!ht]
\centering
\scalebox{0.28}
{\hspace*{0.0cm} \includegraphics[trim={5.0 0.4cm 5.0 0.4cm},clip]{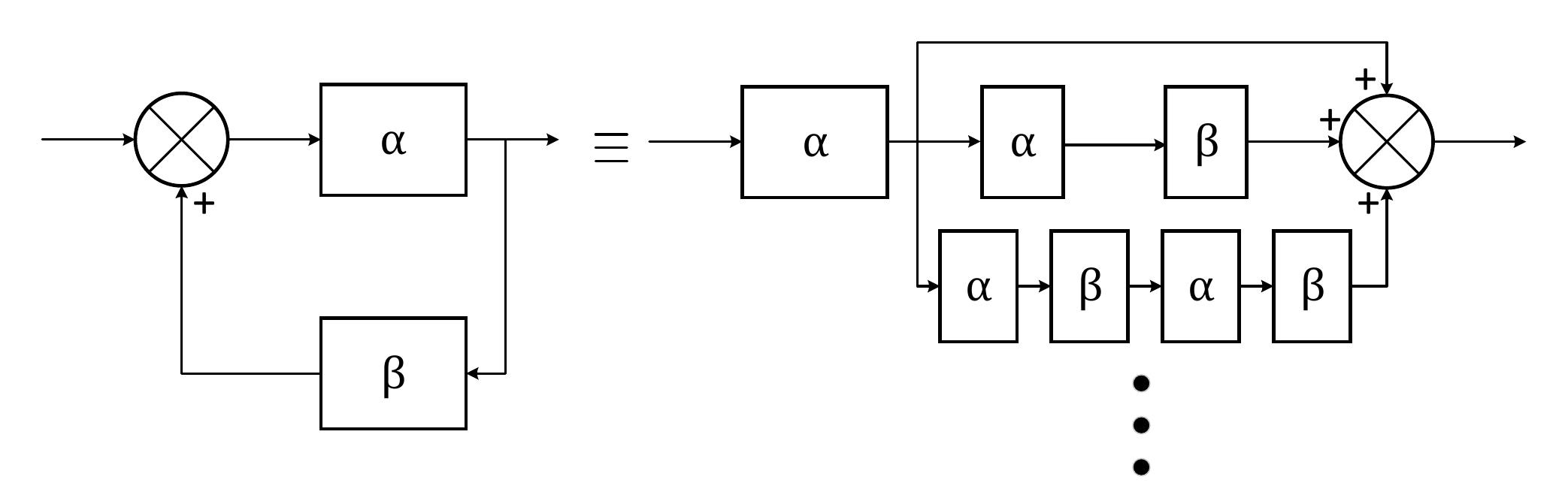}}
\caption{Feedback}
\label{FIG:feedback_block}
\end{figure}

\textbf{Feedback} The feedback configuration serves as the essential representation for modeling the closed-loop response (Figure~\ref{FIG:feedback_block}). Owing to the inclusion of the feedback signal, this configuration is predominantly expressed as an infinite summation of branches comprising of serially connected components. We formally model this as:

\begin{mdframed}
\begin{definition}
\label{DEF:tf_branch}
\emph{Branch} \\
{
\small
\textup{\texttt{\textsf{
$\vdash_{\mathit{def}}$ $\forall \mathtt{\mathsf{\alpha}}\ \mathtt{\mathsf{\beta}}$ n. \kw{branch} $\mathtt{\mathsf{\alpha}}\ \mathtt{\mathsf{\beta}}$ n =   \\
 \hspace*{4.5cm}  $\mathtt{\mathsf{\prod\limits_{i = 0}^{n}}}$ series\_comp [$\mathtt{\mathsf{\alpha}}; \mathtt{\mathsf{\beta}}$]
}}}}
\end{definition}
\end{mdframed}

The above function returns a complex number representing the transfer function of the $n^{th}$ branch by processing the forward path transfer function $\mathtt{\mathsf{\alpha}}$, the feedback signal, the transfer function $\mathtt{\mathsf{\beta}}$ and the number of the branches (\textit{n}).

Now, just like the other blocks, we formally model the feedback block as follows:

\begin{mdframed}
\begin{definition}
\label{DEF:feedback_block_rep_comp}
\emph{Feedback} \\
{
\small
\textup{\texttt{\textsf{
$\vdash_{\mathit{def}}$ $\forall$$\mathtt{\mathsf{\alpha}}$ $\mathtt{\mathsf{\beta}}$. \kw{feedback\_block} $\mathtt{\mathsf{\alpha}}$ $\mathtt{\mathsf{\beta}}$ =  \\
  \hspace*{3.0cm}  series\_comp [$\mathtt{\mathsf{\alpha}}$; $\sum\limits_{k = 0}^{\infty}$ branch $\mathtt{\mathsf{\alpha}}$ $\mathtt{\mathsf{\beta}}$ k]
}}}}
\end{definition}
\end{mdframed}

The above function accepts the forward ($\mathtt{\mathsf{\alpha}}$)  and the feedback path ($\mathtt{\mathsf{\beta}}$) transfer functions. It calculates the overall transfer function by creating a series network that combines the final forward path transfer function with the summation of all potential infinite branches.


\section{Formal Verification of the Ultrafiltration Dialysis Process} \label{SEC:case_study_1}

Ultrafiltration~\cite{henrich2012principles,daugirdas2012handbook} is a process used in kidney dialysis to remove excess fluid and waste products from the blood, particularly for patients with kidney failure or chronic kidney disease. It mimics the function of the kidneys by filtering blood through a semi-permeable membrane. It allows a passage of fluid across a semipermeable (allowing some substances) membrane that is driven by a pressure gradient between the blood and effluent compartments. Since ultrafiltration ensures a desirable level of body fluid by removing its excessive amount during dialysis so its formal verification is of utmost importance for ensuring the correct functionality.

\begin{figure}[!ht]
\centering
\scalebox{0.550}
{
 \includegraphics[trim={5.0 0.4cm 5.0 0.4cm},clip]{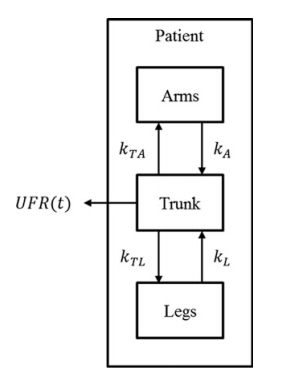}}
\caption{Ultrafiltration Dialysis Process}
\label{FIG:ultrafiltration_dialysis}
\end{figure}

Figure~\ref{FIG:ultrafiltration_dialysis} provides a block (flow) diagram of the ultrafiltration dialysis process for a kidney patient. It is composed of three modules, such as arms, trunk and legs. The volumes of fluid in each of the modules arms, trunk and legs are represented by $V_A(t)$, $V_T(t)$ and $V_L(t)$, respectively. Moreover, the quantity of fluid transferred between these modules are controlled by the transfer constants $k_{TA}$ (between trunk and arms), $k_{TL}$ (between trunk and legs), $k_A$ (between arms and trunk) and $k_L$ (between legs and trunk). The quantity of the fluid removed from the trunk is represented by $UFR(t)$.

The dynamical model of the fluid transfer between arms and trunk is mathematically expressed as:

\begin{equation} \label{EQ:DM_FT_ARMS_TRUNK}
\frac{dV_A(t)}{dt} = -k_A V_A(t) + k_{TA} V_T (t)
\end{equation}

In order to model the dynamics of the fluid transfer, we formalize the generic linear differential equation as follows:

\begin{mdframed}
\begin{definition}
\label{DEF:diff_eq_order_n}
\emph{Linear Differential Equation of Order $n$} \\
{
\small
\textup{\texttt{
$\vdash_{\mathit{def}}$ $\forall$k f t. \kw{diff\_eq\_nord} n lst f t =    \\
\hspace*{0.5cm} vsum (0..n) ($\lambda$i. EL i [$\mathtt{\alpha_\texttt{1}}$;$\mathtt{\alpha_\texttt{2}}$;...;$\mathtt{\alpha_\texttt{i}}$] $\ast$  \\
  \hspace*{2.0cm}  higher\_order\_derivative i f t)
}}}
\end{definition}
\end{mdframed}

The above function takes the order of the diff. equation \texttt{n}, the coefficients list \texttt{l}, a function \texttt{f} and its variable of differentiation \texttt{t}. It outputs the $n^{th}$ order differential equation.

Now, we formalize the dynamics of fluid transfer (Equation~(\ref{EQ:DM_FT_ARMS_TRUNK})) as follows:

\begin{mdframed}
\begin{definition}
\label{DEF:DM_FT_ARMS_TRUNK}
\emph{Dynamics of the Fluid Transfer between Arms and Trunk} \\
{
\small
\textup{\texttt{
$\vdash_{\mathit{def}}$ $\forall$kA. \kw{olst\_diff\_eq\_at} kA = [-- Cx kA; Cx (\&1)]    \\
$\vdash_{\mathit{def}}$ $\forall$kTA. \kw{ilst\_diff\_eq\_at} kTA = [Cx kTA]    \\
$\vdash_{\mathit{def}}$  \kw{diff\_eq\_at} VT VA t kA kTA $\Leftrightarrow$  \\
\hspace*{0.2cm}  diff\_eq\_nord 1 (olst\_diff\_eq\_at kA) VA t =   \\
\hspace*{0.2cm}  diff\_eq\_nord 0 (ilst\_diff\_eq\_at kTA) VT t
}}}
\end{definition}
\end{mdframed}

We now model the corresponding block diagram representation using our formalization in \holl~as:

\begin{mdframed}
\begin{definition}
\label{DEF:block_diagram_FT_ARMS_TRUNK}
\emph{Block Diagram Representation of Fluid Transfer between Arms and Trunk} \\
{
\small
\textup{\texttt{
$\vdash_{\mathit{def}}$ $\forall$kA kTA. \\
\hspace*{1.5cm}  \kw{blk\_diag\_rep\_at} kA kTA = ser $\left[\mathtt{Cx}\ \texttt{kTA};\ \mathtt{\dfrac{Cx (\&1)}{s\ +\ Cx\ kA}}\right]$
}}}
\end{definition}
\end{mdframed}

Next we verify the overall transfer function for this system as follows:

\begin{mdframed}
\begin{theorem}
\label{THM:block_diagram_imp_tf_FT_ARMS_TRUNK}
\emph{Fluid Transfer between Arms and Trunk} \\
{
\small
\textup{\texttt{
$\vdash_{\mathit{thm}}$ $\forall$kA kTA s. \kw{[A]:} s + Cx kA $\neq$ Cx (\&0)     \\
 \hspace*{1.5cm} $\Rightarrow$ blk\_diag\_rep\_at kA kTA = $\mathtt{\dfrac{Cx \&1}{s\ +\ Cx\ kA}}$     \\
}}}
\end{theorem}
\end{mdframed}

We now formally verify the transfer function derived from Theorem~\ref{THM:block_diagram_imp_tf_FT_ARMS_TRUNK}, grounded in the dynamical model, as follows:

\begin{mdframed}
\begin{theorem}
\label{THM:dyn_model_imp_tf_FT_ARMS_TRUNK}
\emph{Model Implies Transfer Function} \\
{
\small
\textup{\texttt{
$\vdash_{\mathit{thm}}$ $\forall$kA kTA VT VA s.    \\
 \hspace*{0.0cm} \kw{A1:} \&0 $<$ kA $\wedge$  \kw{A2:} \&0 $<$ kTA $\wedge$   \\
  \hspace*{0.0cm}  \kw{A3:} $\forall$t.\ diff\_hderivative VT VA t $\wedge$    \\
 \hspace*{0.0cm}  \kw{A4:}   lexst\_hderivative VT VA $\wedge$ \\
 \hspace*{0.0cm}   \kw{A5:}   zero\_init\_conditions VT VA $\wedge$    \\
 \hspace*{0.0cm}  \kw{A6:}  ($\forall$t.\ diff\_eq\_at VT VA t kA kTA) $\wedge$    \\
 \hspace*{0.0cm}  \kw{A7:}  (ltfm VT s $\neq$ Cx (\&0)) $\wedge$  \\
 \hspace*{0.0cm} $\mathbf{\mathtt{\mathsf{\kw{[A_8]:}}}}$  $\mathtt{\mathsf{\left(\dfrac{Cx (\&1)}{s\ +\ Cx\ kA}\ \neq\ Cx (\&0)\right)}}$
  \vspace*{0.1cm} \\
 \hspace*{1.0cm} $\Rightarrow$ $\mathtt{\dfrac{ltfm\ VA\ s}{ltfm\ VT\ s}}$ = $\mathtt{\dfrac{Cx (\&1)}{s\ +\ Cx\ kA}}$
}}}
\end{theorem}
\end{mdframed}

Assumptions \texttt{A1-A2} capture the positivity constraints on the transfer constants \texttt{kTA} and \texttt{kA}. The next 2 assumptions specify the prerequisites for the existance of the first derivative and Laplace transform of \texttt{VA} and \texttt{VT}, respectively. Assumption \texttt{A5} establishes the initial initial conditions for both \texttt{\textsf{VA}} and \texttt{\textsf{VT}}. Assumption \texttt{A6} describes the dynamic behavior of fluid transfer between the arms and trunk. The next 2 assumptions validate the denominator of the transfer function presented in the theorem's conclusion. Ultimately, the conclusion presents the transfer function for the fluid transfer between the arms and trunk.

The theorem-proving-based analysis presented in this work offers significant advantages, making it a rigorous and reliable method for analyzing complex systems like ultrafiltration dialysis. The main strength of the proposed methodology is its ability to explicitly specify all conditions underlying the analysis. Without these assumptions, it would be impossible to validate the correctness of a theorem. This strict requirement guarantees that every proven theorem is sound, making the analysis highly robust and trustworthy. Additionally, the theorems are universally quantified, meaning they are generalizable and can be specialized for any set of values. This flexibility allows the results to be adapted to a wide range of scenarios and configurations, offering great utility in practical applications.

Another key advantage of our approach is its ability to handle the continuities inherent in many real-world systems, such as fluid transfer in ultrafiltration dialysis. Other formal methods, such as SAT-based methods or state exploration based model checking approach, struggle to accurately capture continuous behaviors due to their reliance on discrete-state models. Consequently, theorem proving emerges as a superior formal method for analyzing continuous systems, providing a more accurate representation of the system dynamics.

However, one drawback of this method is the human-interactive nature of the proof process. Theorem proving often requires significant manual intervention, which can be time-consuming and cumbersome, particularly for complex proofs. While this human interaction ensures rigor and flexibility, it also introduces a barrier to efficiency. Fortunately, tactics can be developed to automate certain parts of the proof process, as demonstrated by the automatic tactic \texttt{\textsf{\small{TF\_TAC\_UF}}} used in this work. These tactics can streamline some of the more repetitive or standard aspects of the proofs, reducing the burden on the user while maintaining the overall integrity and soundness of the analysis.

Thus, the proposed theorem-proving approach offers a powerful and reliable method for analyzing biomedical systems, providing both rigor and flexibility unmatched by other formal methods, though it requires ongoing efforts to improve automation.

\section*{Conclusions}
We provide a formal modelling of block diagram representations of biological circuits, specifically for the analysis of biomedical control systems in pHRI. By utilizing higher-order logic (HOL) theorem proving, we have successfully formalized the components and interactions within these systems, providing a systematic and rigorous method for verifying their behavior. This approach is particularly valuable for systems involving complex physiological processes, as it allows for precise analysis and ensures correctness in critical applications.

One key application of this methodology is demonstrated in the formal analysis of the ultrafiltration dialysis process. By modeling the fluid dynamics and transfer functions between different body compartments using block diagrams, we provided a comprehensive framework to verify the system's functionality. The formalization of the dialysis process showcases the practical importance of our approach, ensuring the correct functionality of this life-saving procedure.




\bibliographystyle{unsrt}
\bibliography{bibliotex}

\begin{thebibliography}{10}

\bibitem{fernandez2018automatic}
J~Fern{\'a}ndez, C~Galindo, J~Barbacho, and A~Luque.
\newblock {Automatic Control Systems in Biomedical Engineering}, 2018.

\bibitem{hasan2015formal}
O.~Hasan and S.~Tahar.
\newblock {Formal Verification Methods}.
\newblock In {\em Encyclopedia of Information Science and Technology, Third
  Edition}, pages 7162--7170. IGI Global, 2015.

\bibitem{beillahi2015formal}
S.~M. Beillahi, U.~Siddique, and S.~Tahar.
\newblock {Formal Analysis of Power Electronic Systems}.
\newblock In {\em International Conference on Formal Engineering Methods},
  pages 270--286. Springer, 2015.

\bibitem{siddique2013formal}
U.~Siddique, V.~Aravantinos, and S.~Tahar.
\newblock {Formal Stability Analysis of Optical Resonators}.
\newblock In {\em NASA Formal Methods}, pages 368--382. Springer, 2013.

\bibitem{rashid2018formal}
A.~Rashid, U.~Siddique, and O.~Hasan.
\newblock {Formal Verification of Platoon Control Strategies}.
\newblock In {\em Software Engineering and Formal Methods}, pages 223--238.
  Springer, 2018.

\bibitem{taqdees2017formally}
S.~H. Taqdees and O.~Hasan.
\newblock {Formally Verifying Transfer Functions of Linear Analog Circuits}.
\newblock {\em IEEE Design \& Test}, 34(5):30--37, 2017.

\bibitem{rashid2017formal}
A.~Rashid and O.~Hasan.
\newblock {Formal Analysis of Linear Control Systems using Theorem Proving}.
\newblock In {\em Int. Conf. on Formal Engineering Methods}, volume 10610 of
  {\em LNCS}, pages 345--361. Springer, 2017.

\bibitem{adnanimage}
A.~Rashid, S.~Abed, and O.~Hasan.
\newblock {Formal Analysis of 2D Image Processing Filters using
  Higher-order-logic Theorem Proving}.
\newblock {\em Journal on Advances in Signal Processing}, 53, 2022.

\bibitem{adnancyber}
A.~Rashid and O.~Hasan.
\newblock {Formal Analysis of the Continuous Dynamics of Cyber-physical Systems
  using Theorem Proving}.
\newblock {\em Journal of System Architecture}, 112, 2021.

\bibitem{rashid2020syntheticbiology}
S.~Abed, A.~Rashid, and O.~Hasan.
\newblock {Formal Reasoning about Synthetic Biology using Higher-order-logic
  Theorem Proving}.
\newblock {\em Systems Biology}, 14(5):271--283, 2020.

\bibitem{Ocaml_ref}
M.~Abrams and et~al.
\newblock {{A History of OCaml}}.
\newblock \url{http://ocaml.org/learn/history.html}, 2015.

\bibitem{harrison2006towards}
J.~Harrison.
\newblock {Towards self-verification of HOL Light}.
\newblock In {\em Int. Joint Conf. on Automated Reasoning}, pages 177--191.
  Springer, 2006.

\bibitem{adnan2018JAL}
A.~Rashid and O.~Hasan.
\newblock {Formalization of Lerch’s Theorem using HOL Light}.
\newblock {\em Journal of Applied Logics—IFCoLog Journal of Logics and their
  Applications}, 5(8):1623--1652, 2018.

\bibitem{taqdees2013formalization}
S.~H. Taqdees and O.~Hasan.
\newblock Formalization of {L}aplace {T}ransform {U}sing the {M}ultivariable
  {C}alculus {T}heory of {HOL}-{L}ight.
\newblock In {\em Logic for Programming, Artificial Intelligence, and
  Reasoning}, volume 8312 of {\em LNCS}, pages 744--758. Springer, 2013.

\bibitem{adnan2018TC}
A.~Rashid.
\newblock {{Formal Analysis of the Continuous Dynamics of Cyber-physical
  Systems using Theorem Proving}}.
\newblock \url{http://save.seecs.nust.edu.pk/projects/facdcpstp/}, 2018.

\bibitem{houpis2013linear}
C.~H. Houpis and S.~N. Sheldon.
\newblock {\em {Linear Control System Analysis and Design with MATLAB}}.
\newblock CRC Press, 2013.

\bibitem{henrich2012principles}
W.~L Henrich.
\newblock {\em {Principles and Practice of Dialysis}}.
\newblock Lippincott Williams \& Wilkins, 2012.

\bibitem{daugirdas2012handbook}
J.~T Daugirdas, P.~G Blake, and Todd S.
\newblock {\em {Handbook of Dialysis}}.
\newblock Lippincott Williams \& Wilkins, 2012.

\end{thebibliography}

\end{document}